\documentclass[12pt]{iopart}
\begin{document}
\input{epsf.tex}
\epsfverbosetrue

\title{Dark Matter-Wave Solitons in Optical Lattices}

\author{Pearl J.Y. Louis, Elena A. Ostrovskaya, and Yuri S. Kivshar}

\address{ARC Centre for Quantum-Atom Optics and Nonlinear Physics Group, Research School of Physical Sciences and Engineering, The Australian National University,
Canberra ACT 0200, Australia}

\begin{abstract}
We analyze the Floquet-Bloch matter-wave spectrum of Bose-Einstein
condensates loaded into single-periodic optical lattices and double-periodic superlattices. In the framework of the Gross-Pitaevskii equation, we describe the structure and analyze the mobility properties of dark matter-wave solitons residing on the background of extended nonlinear Bloch-type states. We demonstrate that interaction between dark solitons can be effectively controlled in optical superlattices.
\end{abstract}


\submitto{\JOB; Topical Issue: Optical Solitons}

\section{\label{secIntro}Introduction}

Bose-Einstein condensates (BECs) loaded into optical lattices show a wide
range of interesting physical properties and complex nonlinear dynamics. Optical lattices can be employed to study many
important physical phenomena of atomic and solid state physics. The properties of coherent macroscopic matter waves in a lattice, such as the Bloch-band
structure~\cite{DenschlagSimsarian}, macroscopic interference
effects~\cite{AndersonKasevich}, Bloch oscillations and
Landau-Zener tunnelling~\cite{CristianiMorsch}, have been explored in a number of experiments.

One of the many advantages in using a macroscopic quantum periodic
system, such as BECs in optical lattices is that the 
effective periodic potential created by a standing light wave can be
easily and precisely manipulated by changing the intensities, polarizations, geometry or frequencies of the interfering laser beams. For
example, the depths of the periodic potential wells induced by an
optical lattice can be controlled by tuning the intensities of the
laser beams. The shape of the potential can also be varied by
creating an optical superlattice with more than one order of periodicity. The novel possibilities for band-gap engineering offered by superlattices have been shown to lead to a variety of phase transitions in the contensate \cite{Burnett}, and can be effectively employed for ultracold atom manipulations, such as patterned loading of atoms into lattice sites ~\cite{PeilPorto}. Most remarkably, periodicity of the optical lattice potential leads to the effective dispersion of the BEC wavepackets being a function of the band structure. Thus by inducing the motion of the lattice  relative to the BEC (i.e. via non-zero detuning of the laser beams), one can manage the dispersion properties of a matter wave ~\cite{EiermannTreutlein,Inguscio}.

Coherent matter waves are inherently nonlinear
due to the presence of interatomic interactions. One of the
common features of nonlinear dispersive systems is the
existence of solitons. {\em Bright solitons} are localized
wavepackets where the effects of dispersion are balanced by
nonlinearity to produce self-trapping. In Bose-Einstein
condensates, bright solitons require very special conditions to
be observed. Without a periodic potential, an attractive nonlinearity and a low number of atoms within each soliton (to prevent condensate collapse) is required ~\cite{Strecker,Khaykovich}. In an optical lattice, bright matter-wave solitons have been predicted to exist, due to the balance of repulsive atomic interactions and negative effective dispersion near a band edge~\cite{ZobayPotting,TrombettoniSmerzi,KonotopSalerno,HilligsoeOberthaler, LouisOstrovskaya,OstrovskayaKivshar,CarusottoEmbriaco,EfremidisChristodoulides}. Bright solitons of repulsive BEC take the form of \textit{gap solitons}, embedded in the spectral gaps of the linear Bloch-wave spectrum.

In the majority of condensates currently created experimentally, the interatomic
interaction is repulsive.  This corresponds to an effectively
defocusing nonlinearity of the matter-wave which can support {\em dark solitons} - localized dips on the condensate density background with a phase gradient across the localized region. Similar to other types of solitons, they can remain dynamically stable due to the
balancing effects of nonlinearity and the (positive)
dispersion. Dark solitons have been created experimentally in repulsive condensates by using a phase-imprinting technique to
apply a sharp phase gradient to a condensate cloud in a magnetic trap~\cite{BurgerBongs,DenschlagSimsarian2}. In the case of BECs loaded into optical
lattices, i.e. with the possibility for dispersion management, dark solitons can be supported in both repulsive (for positive
effective dispersion) and attractive
condensates (for negative effective dispersion). Moreover, dark lattice solitons are expected to be easier to create experimentally than bright gap solitons, as they are not confined to the spectral gaps ~\cite{YulinSkyrabin}, and a phase-imprinting technique can be applied to a nonlinear Bloch-wave background within a spectral band. Alternatively, trains of dark solitons can be created in a periodic potential via Bragg scattering \cite{ScottMartin} .

The theory of dark solitons has been developed extensively for many types of
periodic systems such as discrete atomic chains and waveguide
arrays~\cite{PeyrardKruskal,KevrekidisWeinstein,AblowitzMusslimani,SukhorukovKivshar,FengKneubuhl}.
Applying the concepts of discrete dynamical systems to the physics of the Bose-Einstein condensates in optical lattices, Abdullaev \textit{et
al.}~\cite{AbdullaevBaizakov} studied dark and bright solitons on non-zero backgrounds in a vertical lattice by employing a
discrete mean-field model derived in the tight-binding approximation, i.e. considering a single isolated band of the Bloch-wave spectrum.  In
contrast, Yulin and Skryabin~\cite{YulinSkyrabin} used a single-gap
continuous coupled-mode model in order to examine the stability
and existence of out-of-gap dark and bright solitons. A more general analysis
based on the continuous Gross-Pitaevskii equation with a periodic potential
was presented by Alfimov \textit{et al.}~\cite{AlfimovKonotop} who
showed that for a repulsive condensate in an optical lattice, dark
solitons can exist as stationary localized solutions with
nonvanishing asymptotics.  Alfimov~\textit{et
al.}~\cite{AlfimovKonotop} as well as Konotop and
Salerno~\cite{KonotopSalerno} also found numerically stable dark
solitons for periodic quasi-one-dimensional BEC systems. The weak spectral instability of the dark solitons in the combined optical lattice and a strong harmonic potential, both in the discrete and continuous mean-field models, has been established in \cite{Malomed}. 

In this paper we study the structure and mobility properties of dark solitons in single
and double-periodic optical lattices (superlattices) by employing the full continuous mean-field model. First, we describe the multiple band structure of the condensate
Bloch-wave spectrum and analyze the Bloch waves corresponding to
the band edges of the matter-wave spectrum of a single-periodic optical lattice.  We then extend
this analysis to the case of a double-periodic optical
superlattice and demonstrate that extra mini-gaps appear in
the matter-wave spectrum. 
We show that in-band Bloch states can support
dark solitons, and find numerically continuous families of dark
solitons imprinted onto the nonlinear Bloch waves of a repulsive BEC.
These localized states display some properties found in discrete lattice models, including the existence of the pinning (Peierls-Nabarro) potential \cite{KivsharCampbell} that inhibits the solitons mobility and interactions even in the shallow-well case. Pairs of dark solitons with opposing phase
gradients always experience a repulsive interaction in the lattice-free case \cite{KivsharLuther-Davies}, and we show that an optical superlattice can be used to initiate and effectively control the  interactions of dark lattice solitons.

\section{\label{secModel}Model}

The dynamics of a Bose-Einstein condensate loaded into an optical
lattice can be described in the mean field approximation, by the
Gross-Pitaevskii (GP) equation for the macroscopic condensate
wavefunction $\Psi(x,r,t)$,
\begin{equation}
\label{eq3DGPE} i\hbar\frac{\partial\Psi}{\partial t} =
\left\{-\frac{\hbar^2}{2m}\nabla^{2} + V(x,r) +
g_{3D}|\Psi|^{2}\right\}\Psi
\end{equation}
where $r=(y,z)$, $V(x,r)$ is the
time-independent trapping potential, and $g_{3D}=4\pi\hbar^2a_s/m$ characterizes
the two-body interactions for a condensate with atoms of mass $m$
and s-wave scattering length $a_{s}$. The scattering length
$a_{s}$ is positive for repulsive interactions and negative for
attractive interactions. For the cases examined in this paper, we
use the parameters set by $^{87}$Rb: $m=1.44$x$10^{-25}$~kg and
$a_{s}=5.7$~nm.

\begin{figure}
\setlength{\epsfxsize}{12 cm}
\begin{center}
\epsfbox{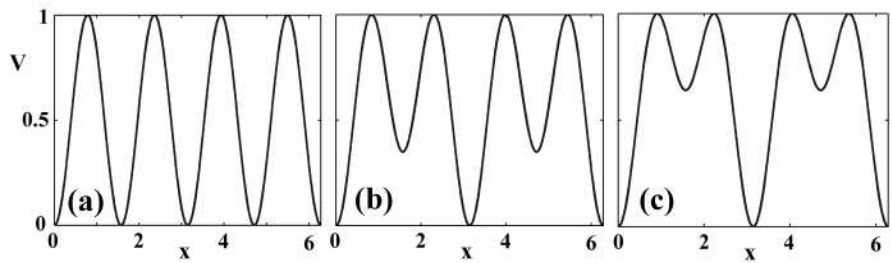}
\end{center}
\caption{\label{figmakingSL} The structure of the superlattice
potential described by Eq. (\ref{pot1DSL}) for different values of
$\varepsilon$ and a fixed potential height $V_0=1$. (a) Single-periodic optical lattice for $\varepsilon=0$, (b,c) double-periodic superlattice with the
height and periodicity unchanged, for (b) $\varepsilon=0.3$ 
 and (c) $\varepsilon=0.5$, respectively.}
\end{figure}

We consider a trapping potential $V(x,r)$ of the form
\begin{equation}
V(x,r) =  \frac{1}{2}m\omega_{\perp}^2(x^2 +
\Omega^2 r^2) + V_{L}(x),
\end{equation}
where $r^{2}=y^2+z^2$, and $\Omega=\omega_\perp/\omega_{x}$.  The first term of the potential describes an anisotropic parabolic potential
due to a magnetic trap, and  $V_{L}(x)$ is the effective
periodic potential formed by a quasi-1D optical lattice of the form,
\begin{equation}
\label{eqPot1}
 V_{L} (x) = U
[\varepsilon\sin^{2}(K_{1}x)+(1-\varepsilon)\sin^{2}(K_{2}x)],
\end{equation}
where $0\leq\varepsilon\leq 1$. The superlattice potential given by
Eq. (\ref{eqPot1}) can be obtained by creating two separate
far-detuned quasi-1D single-periodic lattices using lasers of
different wavelengths, e.g. $\lambda_{1}>\lambda_{2}$.  If the two
lattices are orthogonally polarized, when they are superimposed,
the resulting dipole trapping potential is proportional to the sum
of their individual intensities.  With this interpretation, $U$ is
proportional to the total intensity and $\varepsilon$ related to
the relative intensities of the two standing light waves. The lattice wavevectors are $K_{1}=2\pi/\lambda_{1}$ and
$K_{2}=2\pi/\lambda_{2}$, and the larger of the two periods is $d=\lambda_{1}/2$. In this paper we choose  $\lambda_{1}/\lambda_{2}=2$  (e.g. $\lambda_{1}=700$~nm and $\lambda_{2}=350$~nm)

Eq. (\ref{eq3DGPE}) can be made dimensionless using 
the characteristic length $a_{L}=d/\pi$, energy $E_{rec}=\hbar^{2}/ma_{L}^{2}$, and time $\omega^{-1}_{L}=\hbar/E_{rec}$ scales of the lattice. In dimensionless units, the two-body interaction coefficient is given by
$g_{3D}=4\pi(a_{s}/a_{L})$, and 
the lattice depth is measured in units of the
lattice recoil energy, $E_{rec}$.
Furthermore, assuming that the condensate cloud is strongly
elongated in $x$-direction ($\Omega > 10^{-1}$), the system can be considered to be quasi-1D. The condensate wavefunction
is then separable $\Psi(x,r,t)=\Phi(r)\psi(x,t)$, with
$\Phi(r)$ well described by the ground-state of a
two-dimensional radially symmetric quantum harmonic oscillator,
with the normalization
$\int_{-\infty}^{\infty}|\Phi|^{2}dydz=1$. Integrating dimensionless Eq. (\ref{eq3DGPE}) over the transverse
co-ordinates, gives the 1D GP equation:
\begin{equation}
\label{eq1DGPE} i\frac{\partial\psi}{\partial t} =
\left\{-\frac{1}{2}\frac{\partial^{2}}{\partial x^{2}} + V(x) +
g_{1D}|\psi(x,t)|^{2}\right\}\psi
\end{equation}
where $g_{1D}=8(a_s/a_L)(\omega_L/\omega_\perp)^2$.  As the magnetic confinement along
the axial direction is weak, we can ignore its contribution to
$V(x)$ so that the external potential is approximated by the quasi-1D periodic potential of the optical lattice along the direction of weak confinement:
\begin{equation}
\label{pot1DSL}
V(x)=V_{L}(x)=U[\varepsilon\sin^{2}(x)+(1-\varepsilon)\sin^{2}(2x)],
\end{equation}
where the amplitudes $U$ and $\varepsilon$ are experimentally controlled parameters and can be varied in our model. The shape of an optical superlattice depends on the values of these parameters. In the limits $\varepsilon \to 0,1$, the lattice becomes  single-periodic, and $U$ is equal to the height of the lattice $V_{0}$. For $\varepsilon \neq 0,1$, Eq.(\ref{pot1DSL}) describes a double-periodic superlattice, with the lattice amplitude, $U$, defined through the height of the periodic potential, $V_0$, as
\[
U=16 V_0\frac{1-\varepsilon}{(4-3\varepsilon)^2}
\]  
Figure~\ref{figmakingSL} shows that the superlattice potential consists of a series of alternating small and large wells. As seen in the figure, by changing the intensities of the two component standing waves (i.e. varying $\varepsilon$), the relative depth of lattice wells can be manipulated whilst keeping the height, $V_0$, and the
periodicity of the lattice constant.

\section{The matter-wave band-gap spectrum}
\label{secSpectrum}

The stationary states of a condensate in a quasi-1D infinite periodic
potential are described by solutions of Eq.~(\ref{eq1DGPE}) of the form: $\psi(x,t) = \phi(x) \exp(-i\mu t)$,
where $\mu$ is the corresponding chemical potential. The
steady-state wave function $\phi(x)$ obeys the time-independent GP
equation
\begin{equation}
\label{eq} \left[ \frac{1}{2}\frac{d^{2}}{d x^{2}}
- V(x)+\mu - g_{1D}|\phi(x)|^{2}\right]\phi(x) =0,
\end{equation}
where $V(x)$ is defined in Eq.~(\ref{pot1DSL}).

The case of a noninteracting condensate formally
corresponds to $g_{1D}=0$. Eq.~(\ref{eq}) is then linear in $\phi$ and
the condensate wave function can be presented as a
superposition of Bloch waves,
\begin{equation}\label{BW}
\phi(x) = b_1\phi_1(x) e^{ik x} + b_2\phi_2(x) e^{-ik x},
\end{equation}
where $\phi_{1,2}(x)$ have periodicity of the lattice potential, $b_{1,2}$ are
constants, and $k$ is the Floquet
exponent. The linear matter-wave spectrum
consists of bands of eigenvalues $\mu_{n,k}$ in which $k(\mu)$ is a real wavenumber of amplitude-bounded oscillatory Bloch waves. The bands are separated by ``gaps'' in which $Im(k)\neq0$. The solutions at the band edges are exactly periodic stationary Bloch states.  For an interacting
condensate ($g_{1D}\neq 0$), bright \textit{gap} soliton solutions
exist for the chemical potentials corresponding to the
gaps of the linear matter-wave spectrum~\cite{ZobayPotting,TrombettoniSmerzi,KonotopSalerno,HilligsoeOberthaler,
LouisOstrovskaya,OstrovskayaKivshar,CarusottoEmbriaco,EfremidisChristodoulides}.

\begin{figure}
\setlength{\epsfxsize}{12 cm}
\begin{center}
\epsfbox{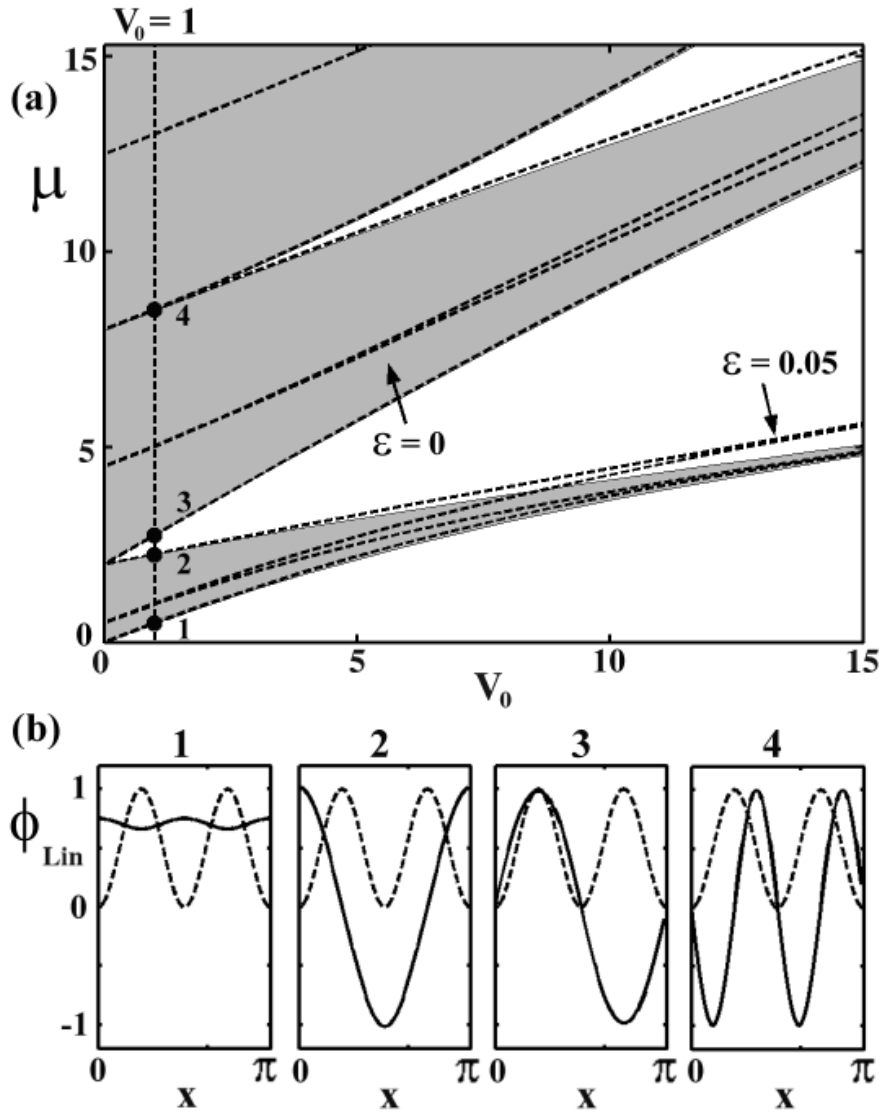}
\end{center}
\caption{\label{figbandsEp0.05} (a) The matter-wave band-gap spectrum of 
Bloch waves in the linear regime (noninteracting condensate). 
The shaded areas show the lowest bands of Bloch matter-waves in a single-periodic
lattice ($\varepsilon=0$). The solid lines mark the band edges
corresponding to periodic Bloch states, examples of which are shown in (b) 
for a lattice of the height $V_{0}=1$. The band edges of the matter-wave spectrum for a 
superlattice at $\varepsilon=0.05$ are shown by the dashed curves.}
\end{figure}

Figure~\ref{figbandsEp0.05} presents the band-gap diagram on the
plane $(\mu, V_0)$ for the Bloch-wave solutions of Eq.~(\ref{eq})
for the case of a noninteracting condensate in the lattice
described by Eq.~(\ref{pot1DSL}) for different lattice parameters. Only the lowest energy transmission bands are shown.  The band-gap structure of a single-periodic optical lattice ($\varepsilon=0$, shaded) is compared with that of a double-periodic lattice or superlattice ($\varepsilon=0.05$, dashed curves). The mini-gaps the superlattice opens up `inside' the transmission
band regions of the single-periodic lattice are clearly visible. As $\varepsilon=0.05$ is
close to $\varepsilon=0$, there is significant overlap between the
two linear band-gap spectra. However, as $\varepsilon\rightarrow
1$, the two spectra diverge. Fig.~\ref{figbandsEp0.05}(b) shows the spatial structure of the Bloch wave solutions, $\phi_{\rm Lin}(x)$ at the edges of the single-periodic transmission bands
for a lattice height of $V_{0}=1$.  Dotted lines show the potential and $\phi_{\rm Lin}$
is normalized according to the condition,
$\int_{0}^{\pi}|\phi_{\rm Lin}|^{2}dx=\pi$.

\begin{figure}
\setlength{\epsfxsize}{12 cm}
\begin{center}
\epsfbox{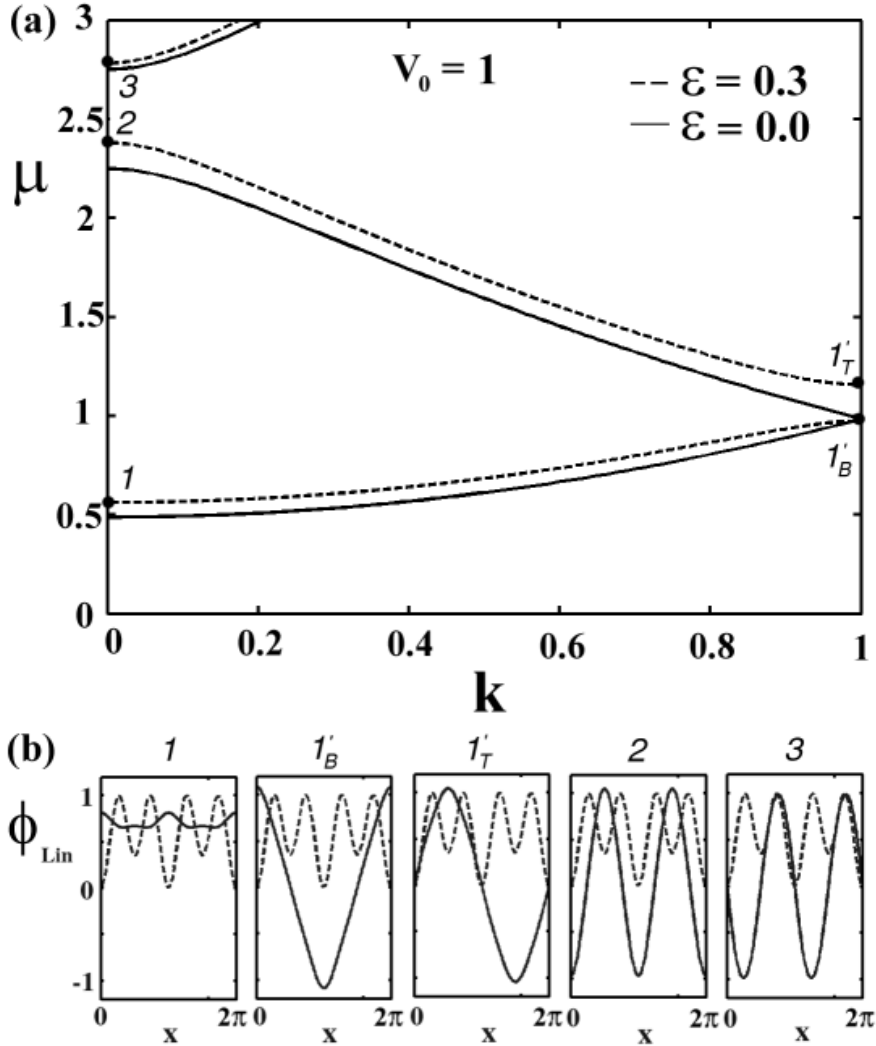}
\end{center}
\caption{\label{figbandsSL} (a) The band-gap spectrum for a noninteracting BEC
in a superlattice (dotted lines: $\varepsilon=0.3$) compared to that of a
single-periodic lattice (solid lines: $\varepsilon=0$), both at $V_{0}=1$. (b) The spatial structure of periodic Bloch-wave solutions at the edges of
the superlattice bands for $\varepsilon=0.3$ and $V_{0}=1$}
\end{figure}

Figure \ref{figbandsSL}(a) shows the band-gap structure for a noninteracting condensate in a superlattice with $\varepsilon=0.3$ compared to that of a
single-periodic lattice, both for a lattice height of $V_{0}=1$.  The
bands of the single-periodic lattice are `folded back' into the first Brillouin
zone ($0 \leq k \leq 1$) of the superlattice. The edges of the first mini-gap
($\mathit{1}_{B}^{'}$, $\mathit{1}_{T}^{'}$) are clearly visible. Fig.~\ref{figbandsSL}(b) shows examples of the structure of the periodic Bloch waves at the edges of the superlattice transmission
bands. The dotted lines show the superlattice potential
and $\phi_{Lin}$ is normalized according to the condition,
$\int_{0}^{2\pi}|\phi_{Lin}|^{2}dx=\pi$.
With the small wells breaking the symmetry of the sinusoidal
potential, Bloch wave solutions exist with peaks in the small wells and/or the
large wells of the superlattice.  The periodic Bloch waves at
band-edges $\mathit{1, 1_{B}^{'}, 1_{T}^{'}}$ and $\mathit{2}$
have discrete counterparts in the solutions of the well-known
diatomic lattice problem. The extended solutions in
higher-order bands  cannot be described using the standard single-band
discrete approximation.

\section{Dark solitons}

\subsection{Single-periodic lattices}

We find different families of dark soliton solutions by using a
standard relaxation technique to numerically solve Eq.~(\ref{eq}) for a lattice potential $V(x)$ given by Eq. (\ref{pot1DSL}).  In this subsection we describe our results for the single-periodic case ($\varepsilon=0$).  In the following subsections we will expand our investigations to the more complicated double-periodic superlattice.  Although we only study
dark solitons associated with the lowest bands of
the matter-wave spectrum, our results should qualitatively apply to dark solitons of the
higher order bands. 

\begin{figure}
\setlength{\epsfxsize}{14 cm}
\begin{center}
\epsfbox{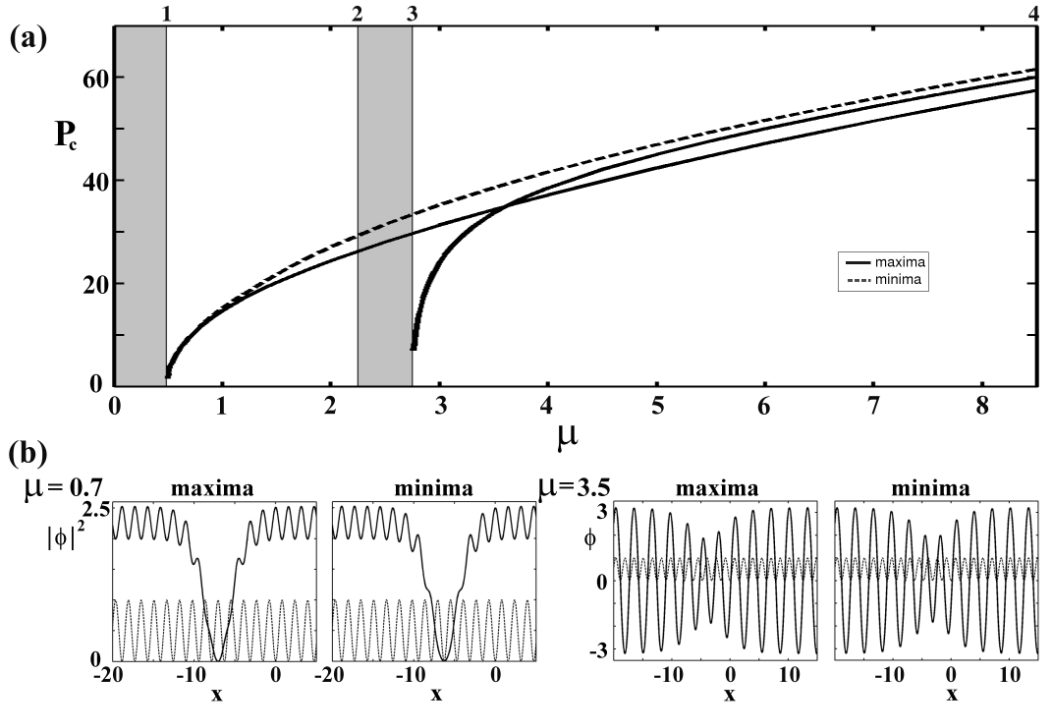}
\end{center}
\caption{\label{figpowerCSingLatt} (a) The complementary powers $P_{c}$
of families of the on-site (solid) and off-site (dashed) dark solitons in a single-periodic lattice
($\varepsilon= 0$) for $V_{0}=1$. Shaded - gaps
of the linear matter-wave spectrum.  Unshaded -
bands, with the band edges numbered. (b) Examples
of the spatial structure of the dark soliton families shown in (a).}
\end{figure}

Figures~\ref{figpowerCSingLatt}(a,b) show the
families of dark solitons for a single-periodic
optical lattice of height
$V_{0}=1$.  According to  Figure~\ref{figbandsEp0.05}, this is in the shallow well regime, where the bands are broad, the condensate behaves
like a superfluid, and the coupling between lattice wells is
strong. Figure~\ref{figpowerCSingLatt}(a) plots the complementary
power, $P_{c}$ of the dark soliton families against the chemical potential
$\mu$. $P_{c}$ is defined as follows,
\begin{equation}
P_{c}=\int \left[ \phi^2_{\mathrm{Background}}(x)-\phi^{2}_{\mathrm{Soliton}}(x) \right]dx
\end{equation}
The complementary power characterizes a deficit of the condensate atoms associated
with the formation of a dark-soliton notch in the Bloch wave background.

By analogy with bright spatially  localized states of the condensate in a lattice, there exist two distinct types of dark solitons - those centered on the minimum (on-site) and maximum (off-site) of the lattice potential. The families of the on-site and off-site dark solitons, characterized by  the dependence $P_c(\mu)$, are plotted in Figure~\ref{figpowerCSingLatt}(a).  The two
families of dark solitons originate at the bottom edge of each band of the
linear matter-wave spectrum. The splitting of the power between the families becomes prominent for large $\mu$ in the first band, but is negligible for the families originating in the second band.
As can be seen in Figure~\ref{figpowerCSingLatt}(b), each of these dark solitons rests
on  a  periodic  matter-wave background. Near the bottom edge of a band, the background
has a spatial structure of a linear periodic
Bloch wave at the corresponding band edge [cf. Fig 2(b)]. For larger $\mu$, when the
linear theory becomes invalid, the background mode describes a
\textit{nonlinear} Bloch wave, which is a periodic solution of the
nonlinear time-independent Gross-Pitaevskii equation~\cite{LouisOstrovskaya}.  The effect
of the interatomic interactions (i.e. nonlinearity) is to
cause an effective shift of the band edges.  Therefore the nonlinear Bloch waves can be
found in spectral regions beyond the linear bands. Correspondingly, the families of dark solitons originate within but extend well beyond the band.

\subsection{Superlattices}

In the previous subsection, we examined the limiting case of
$\varepsilon=0$ corresponding to a single-periodic lattice. Here
we study the dark solitons in a superlattice, where $0<\varepsilon <1$ and a second periodicity is introduced to the system, as shown
in Figure~\ref{figmakingSL} above. As shown in
Section~\ref{secSpectrum}, the addition of the extra periodicity
to the lattice potential causes large structural changes in the
matter-wave band-gap spectrum as well as the spatial structure
of the Bloch states (and hence to the nonlinear Bloch waves).  Dark solitons can now appear at the edges of the mini-gaps opened by the superlattice.

\begin{figure}
\setlength{\epsfxsize}{14 cm}
\begin{center}
\epsfbox{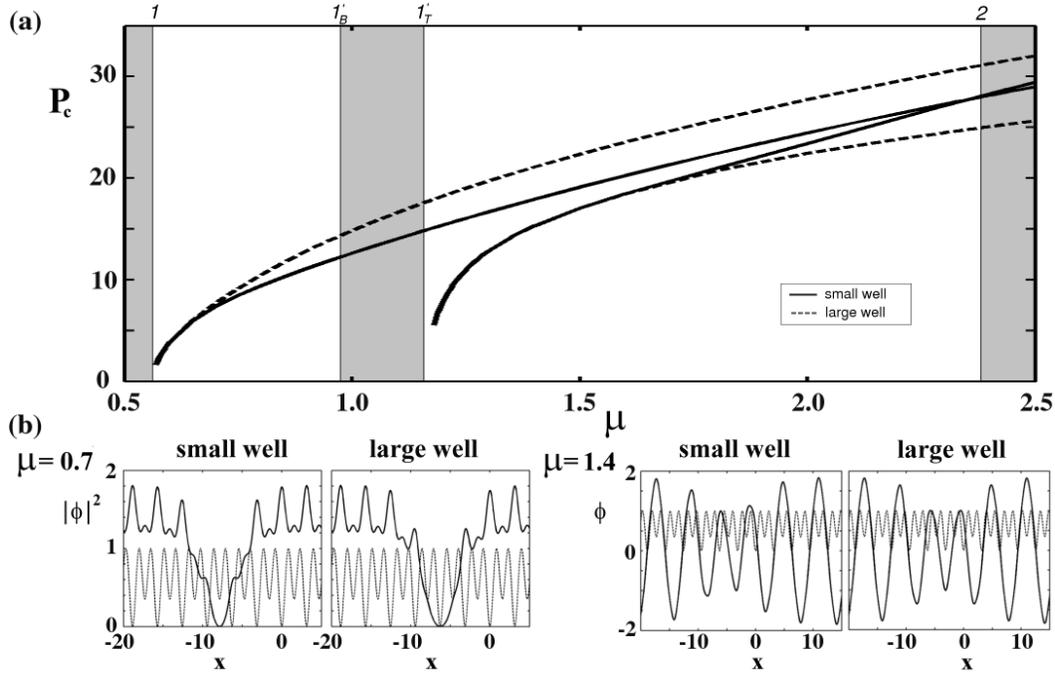}
\end{center}
\caption{\label{figpowerC} (a) The complementary powers $P_{c}$ of two
families of dark solitons in a superlattice ($\varepsilon= 0.3$)
with $V_{0}=1$. Shaded areas show the gap regions of the linear
matter-wave spectrum.  Unshaded areas show the transmission bands
with the band edges numbered. The two bands shown here surround
the first mini-gap (gap edges $1_{B}^{'}$ and $1_{T}^{'}$).
(b) Examples of the spatial structure of the dark soliton families
in (a).}
\end{figure}

We found dark soliton solutions to Eq.~(\ref{eq}) for a
superlattice potential with $\varepsilon=0.3$ and a lattice height of $V_{0}=1$  (shallow well regime) numerically. For this superlattice potential, the minigaps are
inside the bands of the matter-wave spectrum of the single-periodic
lattice.  Figure~\ref{figpowerC}(a) shows the complementary powers of the
dark solitons in the two bands around the first mini-gap opened by
the superlattice potential. The two families in each band correspond to the dark states centered on a large or a small lattice well. The latter state transforms into an off-site dark soliton in the limit $\varepsilon\to1$, and into an on-site soliton  in the limit $\varepsilon\to0$.  Figure~\ref{figpowerC}(b) presents  examples of the spatial structure of each of the dark soliton families.  
As in the case of dark solitons in a single-periodic lattice, the
dark solitons in a superlattice describe kink-like modes on a
non-zero background that, away from the soliton, becomes a nonlinear
Bloch wave.  The families of dark solitons originate at the 
bottom edges of the linear bands and extend beyond the bands as nonlinearity grows.   

As $\mu$ is increased, the power splitting between two families becomes prominent in each of the two bands shown in
Figure~\ref{figpowerC}(a).  In the single-periodic case, the dark solitons centered at a potential minimum always had lower complementary powers than the family centered at the potential
maximum.  A superlattice allows us to have dark solitons centered
in wells of different levels of trapping and as can be seen
Figure~\ref{figpowerC}(a) the more `stable' family seems to
alternate between dark solitons centered at a large well and those
centered at a small well. Similar power splitting behaviour has been observed  for bright gap solitons in binary waveguide arrays recently analyzed in~\cite{PRL_Andrey}; for that model the power splitting can be connected to alternating stability properties. 

Stability is a large and as yet unexplored question for dark
matter-wave solitons in optical lattices.  Several analytical
techniques have been developed to study stability of dark solitons
and are outlined in a review paper by Kivshar and
Luther-Davies~\cite{KivsharLuther-Davies}.  However, a rigorous
analysis of stability of dark matter-wave solitons in an optical
lattice is complicated by the fact that their backgrounds are not constant but take the form of
nonlinear Bloch waves. These have their own stability properties
which have been explored in \cite{Carr,KonotopSalerno}. These studies suggest that only the ground state nonlinear Bloch wave can be modulationally stable.
Some work on the stability of dark matter-wave solitons in
optical lattices have been conducted by Yulin and Skryabin, in the
framework of  a coupled-mode model~\cite{YulinSkyrabin}. Yulin and Skryabin found that
whilst bright out-of-gap solitons on a bright background are
unstable, the stability of dark out-of-gap solitons depends on the
stability of the background~\cite{YulinSkyrabin}. The study of stability of dark
solitons in optical superlattices in the framework of the full
mean-field model is beyond the scope of the present paper.

\subsection{Soliton interactions}

 The difference in the complementary powers of the soliton families described above  can be associated with the difference between the minimal values of the energy functional, 
  \[
 E=\int \left[\frac{1}{2} (\nabla \phi)^2+V_L(x)\phi^2+\frac{1}{2}g_{1D}\phi^4\right] dx,
 \]
 for the two different types of the stationary dark solitons. This energy difference  
corresponds to the maximum of  an effective
potential of the lattice known as the Peierls-Nabarro (PN)
potential~\cite{KivsharCampbell}. The height of the PN potential can be understood as the minimum energy required to move a localized wavepacket by one
lattice site, i.e. the difference between the energies of
a soliton located at a minimum of the periodic lattice and one
at a maximum.  As was shown for bright localized modes in a
discrete lattice, the two different stationary states can be seen
to represent a moving mode at different
times, therefore the knowledge of the PN potential height is essential in answering the questions about possible mobility of a lattice soliton and its ability to interact with other localized states~\cite{KivsharCampbell}.

\begin{figure}
\setlength{\epsfxsize}{11 cm}
\begin{center}
\epsfbox{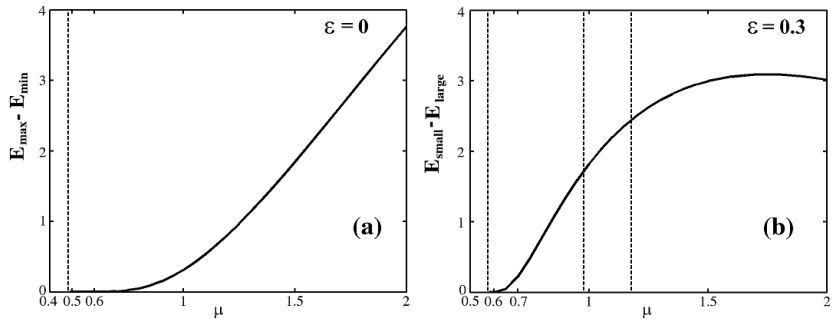}
\end{center}
\caption{\label{pn} PN potential height shown a difference in soliton energies vs. chemical potential for (a) a single-periodic lattice and (b) a superlattice of the same height $V_0=1$. Vertical dashed lines - the band edges.}
\end{figure}

Analytical and numerical calculations of the PN potential for the bright gap solitons have recently been performed in the discrete GP model \cite{Santos}, assuming the tight-binding regime of the BEC dynamics. Remarkably, it has been established that the tight binding calculations of the PN potential fail to match  the results for the full continuous model even in the  regime of parameters where the approximation is well justified. Here, we calculate the PN potential height for the two types of dark solitons found both in the single- and double-periodic lattice, and show the results of the calculations for different values of $\mu$ in Fig. \ref{pn} . In a single-periodic lattice [Fig. \ref{pn} (a)], for the moderate values of $\mu$ within the first band, the dependence has a "flat" region followed by a region where the PN barrier for the on-site dark soliton is positive. Therefore the dark on-site soliton is effectively pinned by the lattice. For the superlattice, the steep positive region of the PN barrier height near the first band edge  [Fig. \ref{pn} (b)] indicates that the motion and interaction of the solitons could be easily initiated, if  initially the dark state is centered on a {\em small well} of the superlattice potential. The inflection of the dependence  at larger $\mu$ could indicate that, for the shallow lattice, the nonlinear effects on the soliton dynamics become more important than those due to the lattice potential. Clearly, for a fixed chemical potential, lattice height, and periodicity, variation of the superlattice parameter $\varepsilon$ controls the height of the PN potential and therefore mobility and interaction properties of the dark solitons.

\begin{figure}
\setlength{\epsfxsize}{10 cm}
\begin{center}
\epsfbox{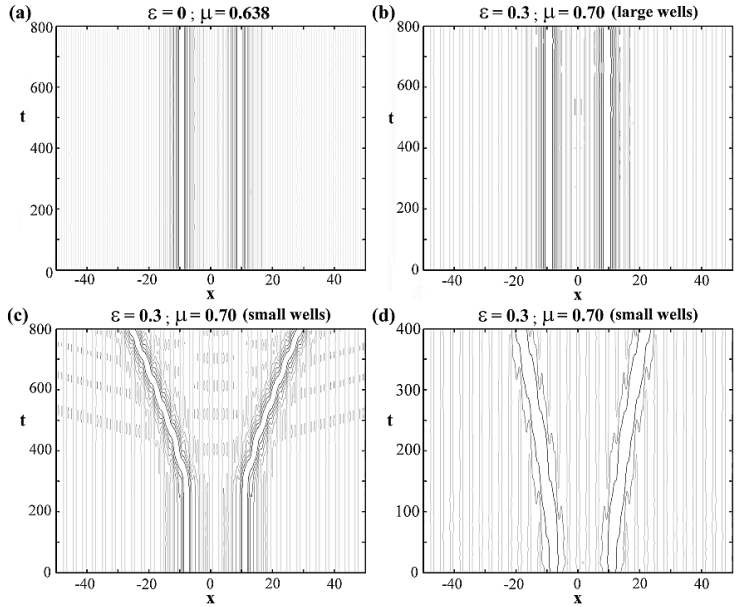}
\end{center}
\caption{\label{figinteractions} Control of the dark soliton
interactions in optical lattices. Shown is the contour plot of the condensate density in the interaction region as a function of time ($V_0=1$). Initially, a pair of dark solitons on a
nonlinear Bloch wave background is centered in (a) the wells of a
single-periodic lattice, (b) the large wells of a superlattice, and (c) the small wells of a superlattice. 
In (d) symmetric amplitude and phase perturbation is added to the initial condition of (c) at $t=0$.}
\end{figure}

To model interactions between dark matter-wave solitons in an
optical lattice we integrate Eq.~(\ref{eq1DGPE}) using a
pseudo-spectral Fourth Order Runge-Kutta method in the interaction
picture.  The code is implemented using the {\bf xmds} code generator~\cite{xmds}.
We study interaction between dark solitons in the framework of two
models of optical lattice potentials described above, i.e. (i) a
single-periodic lattice with $\varepsilon=0$ and (ii) an optical
superlattice with $\varepsilon=0.3$.  In each case, the lattice
height is fixed to $V_{0}=1$. The initial conditions for both
lattices are a pair of dark solitons on a ground-state nonlinear Bloch wave
background [see Figs. 4(b) and 5(b)].
The chemical potential is slightly different in each case so as to
have the same maximum density (and hence nonlinearity) for both
lattices.  The initial spatial separation between the dark
solitons in a pair is $6\pi$.  For the superlattice this
corresponds to a spatial separation of 6 wells.  For the single-
periodic lattice, a spatial separation of 12 wells.  The results of the temporal dynamics
are shown in Figure~\ref{figinteractions} (for the values of $\mu$ near the bottom edge of the first band), and seem to agree with the qualitative predictions from the PN calculations.

Figure~\ref{figinteractions}(a) shows that even in the
shallow-well regime, a single- periodic potential is sufficient to
trap the dark solitons due to the effect of the large positive PN
potential height (both solitons are initially centered at the potential minima). In this case, the density perturbation of the matter wave develops from the numerical noise. By adding the amplitude perturbation to the initial conditions, one can observe damped oscillations of the individual soliton positions around the potential minimum (not shown). In the
superlattice potential, if both solitons are initially centered at the large
wells of the lattice potential, the dark solitons are also trapped
by the lattice, as in Figure~\ref{figinteractions}(b). However, if
the solitons are shifted so that they are initially centered at
the small wells of the lattice potential they are able to move and 
repel each other, as shown in Figure~\ref{figinteractions}(c). The addition of amplitude perturbations to the initial state almost immediately triggers a rapid soliton separation Figure~\ref{figinteractions}(c). 
The same effect can be achieved by changing the contrast of the superlattice by increasing $\varepsilon$ whilst keeping the periodicity and lattice height constant.

\section{Conclusions}

We have analyzed the band-gap structure of the Floquet-Bloch
matter waves in optical lattices and superlattices in the framework of the
Gross-Pitaevskii equation with a single- and double-periodic potential. We have shown
that each type of the nonlinear Bloch matter-waves can support dark solitons - localized, stationary notches in the condensate background density with a phase gradient.  We have described different families of dark solitons originating within the multiple bands of the Floquet-Bloch matter wave spectrum. By considering the continuous analogue of the Peierls-Nabarro potential in discrete lattices, we have demonstrated that the mobility and  interaction properties of the dark solitons can be effectively controlled by changing the structure of an optical superlattice.

\ack The authors acknowledge a support from the Australian
Research Council and useful discussions with Dr. Craig Savage.

\end{document}